\documentclass[12pt]{article}
\usepackage{amssymb,amsmath,epsfig}
\allowdisplaybreaks

\begin{document}
\title{\bf Impact of $f(\mathcal{Q})$ Theory on the Stability of Compact Spherical Solutions}
\author{Shamaila Rani $^{1,2}$ \thanks {shamailatoor.math@yahoo.com;
drshamailarani@cuilahore.edu.pk}, Muhammad Adeel $^2$
\thanks{mr.adimaths@gmail.com},  M. Zeeshan Gul $^{3,4}$
\thanks{mzeeshangul.math@gmail.com}, \\ and Abdul Jawad $^{1,2}$
\thanks
{jawadab181@yahoo.com; abduljawad@cuilahore.edu.pk}
$^1$~Institute for Theoretical Physics and Cosmology,\\
Zhejiang University of Technology, Hangzhou 310023, China\\
$^2$ Department of Mathematics, COMSATS University Islamabad,\\
Lahore-Campus, Lahore-54000, Pakistan.\\
$^3$ Department of Mathematics and Statistics, The University\\ of
Lahore 54792, Pakistan.\\
$^4$ Research Center of Astrophysics and Cosmology, Khazar
University, \\ Baku, AZ1096, 41 Mehseti Street, Azerbaijan.}

\date{}

\maketitle

\begin{abstract}
This research paper examines the feasibility and stability of
compact stars in the context of $f(\mathcal{Q})$ theory, where
$\mathcal{Q}$ represents the non-metricity scalar. To achieve this
objective, a static spherical line element is assumed in the
interior region and the Schwarzschild spacetime is used in the
exterior region of the star. The unknown constants are determined by
using the Darmois junction conditions. We consider a specific model
of this theory to investigate the viability of compact stars through
various physical quantities such as matter contents, energy bounds,
anisotropy and state parameters. The stability states for the
stellar objects under consideration are determined by the speed of
sound and adiabatic index, respectively. The resulting data indicate
that the compact stars in this modified framework are physically
viable and stable.
\end{abstract}

\section{Introduction}

The recent advancements in gravitational physics and cosmic science
have brought significant attention to the rapid expansion of the
universe \cite{1}-\cite{2}. These developments have provided new
insights into understanding the fundamental and practical
modifications that contribute to the rapid growth of galaxies. Many
results have provided strong evidence to understand the expansion of
the universe \cite{3}-\cite{5}. The universe is expanding faster
than ever before due to an unknown factor known as dark energy.
Moreover, it is estimated that 68 percent of total energy is
unexplained in the form of DE. Therefore, researchers need to make
some modifications to the general theory of relativity (GR) to
determine the cosmic expansion. These kinds of studies encourage
researchers to investigate viable modifications or extensions of
gravity theories that could simulate the situations when the GR
yields unfavorable results. Cosmologists have explored some
alternative gravitational theories in response to the limitations of
GR. There are different forms of modified theories such as
curvature, torsion and non-metricity-based theories
\cite{4a}-\cite{5j}.

Modified theories of gravity successfully offer the alternative
explanations to GR by addressing unsolved issues of the universe.
Recently, extensive research has delved into the implications of
modified theories. Weyl in his work \cite{21} integrated
electromagnetic and gravitational forces into a more generic
geometry by introducing the concept of non-metricity (covariant
divergence is zero for metric tensor). Cartan \cite{23} developed
the Einstein-Cartan theory by extending GR in the presence of
torsion field. Weitzenbock space \cite{24} features zero curvature
and torsion replaces curvature in teleparallel theory. Introduction
of non-metric variables such as the non-metricity in symmetric
teleparallel gravity, further refines this framework. Notably, both
torsion and curvature remain zero in this formulation.

The $f(\mathcal{Q})$ theory explores an alternative route where
non-metricity is the driving factor, capturing considerable
attention among researchers due to its profound implications in
gravitational physics. The exploration of this theory is motivated
by a desire to understand its theoretical implications and
significance in astrophysical as well as cosmological contexts.  In
$f(\mathcal{Q})$ theory, the inclusion of non-metricity enables a
more detailed depiction of gravitational interactions. Non-metricity
represents the deviation from the Levi-Civita connection, which is
the connection compatible with the metric tensor in GR. By
incorporating non-metricity into the gravitational action, this
theory introduces additional degrees of freedom, leading to new
gravitational dynamics and cosmological solutions. This modification
influences the internal structure of pulsars, leading to alterations
in the pressure-density relationship and changes in stellar radii
and mass profiles. Consequently, the equations of motion derived
from this theory impact the hydrostatic equilibrium and stability of
stars. Neutron stars serve as prime sources of gravitational waves
which exhibit distinct signatures under the influence of
$f(\mathcal{Q})$ gravity, differing from those predicted by GR.
Noteworthy, outcomes highlight the presence of additional correction
terms, potentially influencing the geometric viability when compared
to GR. The exploration of this theory is motivated by a desire to
understand its theoretical implications and significance in
astrophysical as well as cosmological contexts.

The non-metricity is a mathematical concept that emerges in theories
involving non-Riemannian geometries, providing an alternative cosmic
model without dark energy. Researchers are drawn to explore
non-Riemannian geometry, specifically $f(\mathcal{Q})$ theory, for
various reasons such as its theoretical implications, compatibility
with observational data and its significance in cosmological
contexts \cite{24a}. Recent investigations into $f(\mathcal{Q})$
gravity have revealed cosmic issues and observational constraints
can be employed to indicate deviations from the $\Lambda CDM$ model
\cite{24b}. Spherical symmetric configurations in $f(\mathcal{Q})$
gravity have been analyzed in \cite{24c}. Ambrosio \cite{24d}
described perturbation corrections to the Schwarzschild solution in
the same theory. Ambrosio et al \cite{24e} delved into the
asymptotic behavior of Schwarzschild-like solutions in
$f(\mathcal{Q})$ theory. The non-metricity scalar has been employed
to detect the effects of microscopic systems in \cite{24f}. The
viable cosmological solutions in symmetric teleparallel gravity
through the Noether symmetry technique have been explored in
\cite{24g}. Barros et al \cite{24h} analyzed the cosmic
characteristics through redshift space distortion data in
non-metricity gravity. This modified theory can elucidate the cosmic
bounce scenario \cite{24i} and describes dark energy features at
large scales \cite{24j}. For further details, we refer the readers
to \cite{24k}-\cite{24p}.

Adak \cite{25a} studied the symmetric teleparallel gravity model in
which only non-metricity is non-zero. They obtained a spherically
symmetric static solution to Einstein equation in symmetric
teleparallel gravity and discussed the singularities. Nester and Yo
\cite{25b} studied teleparallel geometry with zero curvature and
torsion while non-zero nonmetricity behaves as the gravitational
force. Adak and Sert \cite{25c} explored a gravity model that is
characterized by nonmetricity, discovering that the horizon becomes
singular in symmetric teleparallel gravity. Adak et al \cite{25d}
formulated a symmetric teleparallel gravity model incorporating the
Lagrangian in the non-metricity tensor, comprehensively analyzing
the variations applicable to gravitational formulations. They
derived a set of solutions encompassing Schwarzschild,
Schwarzschild-de Sitter and Reissner-Nordstrom solutions for
specific parametric values. The spherical symmetric configuration in
$f(\mathcal{Q})$ gravity was investigated in \cite{25e}. Maurya et
al \cite{25f} noted the significant impact of the nonmetricity
parameter and decoupling constant on the stability of CSs in
$f(\mathcal{Q})$ gravity. Adak et al \cite{25g} delved into the
broader realm of teleparallel geometry using differential forms.
Their exploration encompassed the examination of specific instances
such as metric and symmetric teleparallelism. They provided insights
into the connections between formulations employing gauge fixings
and those without gauge fixing. Additionally, the researchers
introduced a technique for transforming Riemannian geometries into
teleparallel structures.

Cosmological observations revealed that the universe is originated
through a significant expansion of matter and energy. Scientists
raise sparking numerous inquiries among the formation, evolution and
structure of the universe. Delving into the enigmatic features of
the universe cosmologists have focused on the roles played by
galaxies, planets, clusters and stars. Cosmology also elucidates
fundamental processes such as the evolution and ultimate fate of
stars. In maintaining its equilibrium, star relay on the delicate
balance between inward gravitational force and outward matter
pressure. This equilibrium becomes precarious when pressure is
insufficient to counterbalance the gravity, resulting in collapse of
the star. This collapse gives rise to new compact objects with
reduced radii known as compact remnants. Many researchers have been
intrigued by the nature and precise composition of compact stars
(CSs).

Among these celestial bodies, neutron stars have garnered
significant interest due to their captivating properties and
structures. Baade and Zwicky \cite{26} delved into the geometry of
CSs attributing their formation to supernovae and confirming their
existence with the discovery of pulsars \cite{27}. The presence of
neutron stars becomes valid by identification of pulsars which
captivate the researchers to analysis its various attributes. Dev
and Gleiser \cite{28} explored physical attributes of rotating
neutron stars. Mak and Harko \cite{29} scrutinized the physical
characteristics of pulsars assessing their stability based on sound
speed method. Kalam et al \cite{30} examined viable and stable CSs
using Krori-Barua solutions. Rahaman et al \cite{31} investigated
the physical characteristics of CSs by employing EoS parameters.
Maurya et al \cite{32} explore viable characteristics of anisotropic
stars by considering different metric elements. Singh and Pant
\cite{33} applied a similar technique to ascertain the interior
geometry of CSs. Bhar et al \cite{34} assessed viable and stable
CSs. Jasim et al \cite{35} delve a non-singular spherical symmetric
for anisotropic weird stars using the Tolman-Kuchowicz solution.

Isotropy (equal principal stresses) is a common assumption in the
study of self-gravitating systems, whenever the fluid approximation
is used to describe the matter distribution of the object. This
character of fluids is supported by a large amount of observational
evidence, pointing towards the equality of principal stresses under
a variety of circumstances. However, strong theoretical evidences
presented in the last decades suggest that for certain density
ranges, different kinds of physical phenomena may take place, giving
rise to anisotropy. The number of physical processes giving rise to
deviations from isotropy is quite large in both high and low density
regimes. Thus, for highly dense systems, exotic phase transitions
may occur during the process of gravitational collapse. The
consideration of anisotropic fluids in the context of CSs is
essential to accurate capture the complex physical phenomena that
govern their structure and stability. The inclusion of anisotropic
pressure allows for a more realistic representation of matter under
extreme conditions, leading to improved theoretical models that
align better with observational data and predictions regarding
stellar evolution, gravitational collapse and oscillation modes. In
the extreme gravitational environments of CSs, the pressure and
density may not be uniform throughout the star.

Anisotropic fluids account for the directional dependence of
pressure, which can arise from several factors such as nuclear
interactions, rotation and magnetic fields. At high densities, the
interactions between particles can create different pressures in
different directions due to the influence of strong nuclear forces.
Rapid rotation and strong magnetic fields can induce anisotropic
pressure distributions. For instance, a rotating star may experience
greater pressure along its equatorial plane compared to its poles.
Anisotropic models allow for a more flexible and realistic equation
of state that can better describe the thermodynamic properties of
the matter in these stars. In dense astrophysical environments,
matter may undergo phase transitions (e.g., from hadronic matter to
quark-gluon plasma), leading to anisotropic pressure components.
Anisotropic fluids can lead to different stability conditions
compared to isotropic ones. In CS models, stability against
gravitational collapse and oscillation modes can be better
understood with anisotropic pressures, which might stabilize certain
configurations that would be unstable in isotropic models.
Anisotropic pressure can play a crucial role in avoiding
singularities within the star, allowing for solutions that are
physically viable. The detection of gravitational waves from neutron
star mergers has revealed complex dynamics that can be modeled more
accurately with anisotropic fluid dynamics, reflecting the actual
physical processes at play. Anisotropic models can provide a richer
structure for the field equations, enabling solutions that capture
the complex nature of CSs. The influence of anisotropy has been
extensively studied in \cite{36}-\cite{50}.

In the study of CSs, the assumption of isotropic pressure might be
convenient for simplification, but it often fails to capture the
complex dynamics present in these objects. The physical processes
inherent in stellar evolution, including dissipative phenomena and
dynamic equilibrium, inherently lead to the development of pressure
anisotropy, reinforcing the necessity to consider it in our models.
Incorporating pressure anisotropy into the modeling of CSs is not
merely an academic exercise, but it reflects the underlying physical
realities of stellar dynamics. The inherent dissipative processes
during stellar evolution guarantee that anisotropic pressures will
manifest, shaping the final equilibrium state. Thus, models that
overlook this essential feature risk failing to accurately describe
the behavior of these fascinating astrophysical objects. Herrera
\cite{51} highlighted an important aspect that the pressure
anisotropy is not merely a consequence of specific physical
conditions in compact objects, but is rather an unavoidable outcome
of the stellar evolution process. The consideration of anisotropic
fluids in our model not only aligns with the physical phenomena
expected in compact objects but also accommodates the natural
evolution of pressure anisotropy during the system history. By
adopting anisotropic fluid models, researchers can better capture
the essential characteristics of CSs, leading to more accurate
predictions of their behavior such as mass-radius relationships and
stability criteria.

In astrophysics, CSs are objects that possess incredibly high
densities and are supported by the balance between gravity and
pressure due to their matter composition. To describe the internal
structure and gravitational properties of such stars, it is crucial
to find suitable solutions to the Einstein field equations that
account for the behavior of matter under extreme gravitational and
pressure conditions. One such solution is the Tolman-Kuchowicz
spacetime. The Tolman-Kuchowicz spacetime is a specific solution to
the field equations that has been applied to model the interior
structure of CSs, such as neutron stars. The motivation for using
this spacetime lies in capturing the essential physical properties
of dense stellar objects and providing a framework for studying the
behavior of matter and spacetime in extreme conditions. The
Tolman-Kuchowicz spacetime is designed to satisfy several physical
conditions necessary for realistic stellar models, i.e., regularity
at the center. The metric components are regular at the center of
the star, avoiding singularities and providing a smooth geometry.
The solution can describe matter distributions with decreasing
density and pressure from the center to the surface, mimicking the
expected physical behavior of CS interiors. Thus, the
Tolman-Kuchowicz spacetime is physically motivated in the context of
CSs as it provides an exact solution to the field equations that
satisfies the necessary physical conditions required for realistic
models of CSs. It allows astrophysicists to explore how extreme
densities and pressures in these stars influence their overall
stability, making it a valuable tool for understanding the nature of
compact objects.

The gravitational decoupling via complete geometric deformation
method that has been introduced to explore the nonmetricity effects
in relativistic astrophysics in \cite{52}. Lohakare et al \cite{53}
studied simulates strange stars in $f(\mathcal{Q})$ gravity with an
additional source under an electric field using gravitational
decoupling technique. Maurya et al \cite{54} examined compact
objects in the framework of gravity by employing the method of
gravitational decoupling. Kaur et al \cite{55} presented a study on
a spherically symmetric anisotropic solution in $f(\mathcal{Q})$
gravity in the framework of vanishing complexity formalism. Maurya
et al \cite{56} attempted to find an anisotropic solution for a CS
generated by gravitational decoupling in $f(\mathcal{Q})$ gravity
theory having a null complexity factor. Maurya et al \cite{57}
investigated the effect of $f(\mathcal{Q})$ parameters on two
classes of exact spherically symmetric self-bound isotropic
solutions for compact objects. Kumar et al \cite{58} investigated
the physical behavior and stability of charged CSs in
$f(\mathcal{Q})$ gravity. Errehymy \cite{59} et al presented a
rigorous study of compact objects beyond the standard limit in
$f(\mathcal{Q})$ gravity in particular $f(\mathcal{Q})=b_1
\mathcal{Q}+b_2$, where $b_1$ and $b_2$ are constants. Chaudhary et
al \cite{60} examined the properties of wormhole solutions in the
this modified context. They analyzed that their findings contribute
to the current discussion about alternative gravity theories and
exotic spacetime geometries. Maurya et al \cite{61} achieved a
significant leap forward in understanding compact stellar systems
and the modified $f(\mathcal{Q})$ gravity theory. Their
investigation advances our knowledge of compact stellar systems and
supports the evolution of the modified $f(\mathcal{Q})$ theory of
gravity, opening the way for more breakthroughs in this field.
Maurya et al \cite{62} explored the possibility of existing a novel
class of compact charged spheres based on a charged perfect fluid in
the realm of $f(\mathcal{Q})$ theory. Mustafa et al \cite{63}
predicted the mass-radius relation along with the maximum mass limit
of several objects for different parameter values by assuming two
different surface densities and discovered that the compactness
rises when density increases. We have also discussed the compact
objects in different modified theories of gravity
\cite{63jj1}-\cite{63jj3}.

Jasim et al \cite{64} analyzed a singularity-free model for the
spherically symmetric anisotropic strange stars under Einstein
general theory of relativity by exploiting the Tolman-Kuchowicz
metric. Biswas et al \cite{65} proposed a relativistic model of a
static spherically symmetric anisotropic strange star with the help
of Tolman-Kuchowicz metric potentials. Shamir and Fayyaz \cite{66}
presented the Einstein-Maxwell equations which are described by the
spherically symmetric spacetime in the presence of charge by
exploiting the Tolman-Kuchowicz spacetime. Shamir and Naz \cite{67}
explored some relativistic configurations of stellar objects for
static spherically symmetric structures in the context of modified
gravity by exploiting the Tolman-Kuchowicz spacetime. Raj et
\cite{68} studied the geometry of SAX J 1808.4-3658 CS with charged
matter configuration in the framework of modified gravity theory
using the metric potentials proposed by Tolman-Kuchowicz. Ditta and
Xia \cite{69} explored the stellar structures by employing Karmarkar
condition Tolman-Kuchowiz metric components with the anisotropic
source of the matter distribution in the background of Rastall
teleparallel gravity. Bandyopadhyay and Biswas \cite{70} studied the
mass-radius relation of the specific CS in the background of the
late time accelerated Universe and to achieve the equation of state
of the core matter under stable equilibrium in modified gravity
using Tolman-Kuchowicz metric potentials.

Singh et al \cite{71} examine the physical behavior of CSs through
karmarkar condition in $f(\mathcal{R},\mathrm{T})$ theory. The use
of karmarkar condition to investigate the viable behavior of strange
stars in $f(\mathcal{R},\mathrm{T})$ theory has been studied in
\cite{72}. Biswas et al \cite{73} proposed a realistic model for a
static spherically symmetric anisotropic weird star using
Tolman-Kuchowicz metric potentials. Shamir and Naz \cite{74}
utilized the Tolman-Kuchowicz spacetime to investigate
configurations of static spherically symmetric structures under
altered theory. In Brans-Dicke theory, Majid and Sharif \cite{75}
developed anisotropic model for weird star by deriving field
equations for the Tolman-Kuchowicz ansatz. Javed et al \cite{76}
examined anisotropic sphere star and derived equations of motion for
anisotropic matter distribution. Maurya et al \cite{77} explored
anisotropic matter distributions of CSs in the curvature-matter
coupled theory.

This paper is structured in following pattern. Section \textbf{2}
derives the equations of motion using the Tolman-Kuchowicz solution.
This section also manipulates the unknown parameters through
matching conditions. In Section \textbf{3}, we evaluate the physical
behavior of the CSs. In Section \textbf{4} stability of considered
CSs is influenced through the sound speed and adiabatic index.
Finally, we present a summary of our findings and implications in
section \textbf{5}.

\section{Extended Symmetric Teleparallel Theory}

The corresponding action is expressed as \cite{93}
\begin{equation}\label{1}
S=\int\left(\frac{1}{2}f(\mathcal{Q})+ L_{m}\right)\sqrt{-g}d^4x.
\end{equation}
The non-metricity and deformation tensor are defined as
\begin{eqnarray}\label{1a}
\mathcal{Q}&\equiv&-g^{\varphi\psi}(L^{\varrho}_{~\iota\varphi}
L^{\iota}_{~\varrho\psi}
-L^{\varrho}_{~\iota\varrho}L^{\iota}_{~\varphi\psi}),
\\\nonumber
L^{\varrho}_{~\iota\varphi}&\equiv&-\frac{1}{2}
g^{\varrho\varepsilon}
(\nabla_{\varphi}g_{\iota\varepsilon}+\nabla_{\iota}
g_{\varepsilon\varrho} -\nabla_{\varepsilon}g_{\iota\varphi}).
\end{eqnarray}
The superpotential in terms of non-metricity is given by
\begin{equation}\label{1c}
P^{\varrho}_{~\varphi\psi}=-\frac{1}{2}L ^{\varrho}_{~\varphi\psi}
+\frac{1}{4}(\mathcal{Q}^{\varrho}
-\bar{\mathcal{Q}}^{\varrho})g_{\varphi\psi}- \frac{1}{4} \delta
^{\varrho} _{(\varphi \mathcal{Q}_{\psi})}.
\end{equation}
The relation for $\mathcal{Q}$ using superpotential is expressed as
\begin{equation}\label{1d}
\mathcal{Q}=-\mathcal{Q}_{\varrho\varphi\psi}P
^{\varrho\varphi\psi}=-\frac{1}{4} (-\mathcal{Q}^{\varrho\psi\iota}
\mathcal{Q}_{\varrho\psi\iota}+2\mathcal{Q}^{\varrho\psi\iota}
\mathcal{Q}_{\iota\varrho\psi}
-2\mathcal{Q}^{\iota}\bar{\mathcal{Q}}_{\iota}+\mathcal{Q}
^{\iota}\mathcal{Q}_{\iota}).
\end{equation}
The resulting field equations are
\begin{equation}\label{7}
\frac{2}{\sqrt{-g}} \nabla_{\varrho}(\sqrt{-g}
f_{\mathcal{Q}}P_{\varphi\psi}^{\varrho})+
\frac{1}{2}g_{\varphi\psi} f + f_{\mathcal{Q}}(P_{\varphi \varrho
\epsilon}\mathcal{Q}_{\psi}~^{\varrho
\epsilon}-2\mathcal{Q}_{\varrho \epsilon\varphi}
P^{\varrho\epsilon}~_{\psi}) = -T_{\varphi\psi},
\end{equation}
where $f\equiv f(\mathcal{Q})$ and $f_{\mathcal{Q}}=\frac{\partial
f}{f_{\mathcal{Q}}}$. The solution to these modified field equations
may provide valuable understanding of gravity's behavior in this
framework.

We analyze the compact stellar structures by studying static
spherical spacetime as
\begin{equation}\label{9}
ds^{2}=-dt^{2}e^{\nu(r)}+dr^{2}e^{\lambda(r)}+d\theta^{2}r^{2}+
d\phi^{2}r^{2}\sin^{2}\theta,
\end{equation}
where $e^{\nu(r)}$ and $e^{\lambda(r)}$ are metric potentials which
ensures that the spacetime is static. We consider the anisotropic
matter distribution as
\begin{equation}\label{10}
T_{\varphi\psi}=\rho
u_{\varphi}u_{\psi}+(u_{\varphi}u_{\psi}+g_{\varphi\psi})P_{t}+(P_{r}-P_{t})v_{\varphi}
v_{\psi},
\end{equation}
where $u_{\varphi}$ and $v_{\psi}$ represent four-velocity and
four-vector of the fluid, respectively. Using Eq.(\ref{9}), we have
\begin{equation}\label{11}
\mathcal{Q}=-\frac{2}{r^{2}e^{\lambda}}(1+r \nu'),
\end{equation}
where prime represents the radial coordinate derivative. The
resulting field equations are given as follows
\begin{eqnarray}\label{12}
\rho&=&\frac{f}{2}- f_{\mathcal{Q}}\big(\mathcal{Q} +
\frac{1}{r^{2}}+ \frac{e^{-\lambda}}{r}(\nu'+\lambda')\big),
\\\label{13}
P_{r}&=&f_{\mathcal{Q}}\big(\mathcal{Q}+\frac{1}{r^{2}}\big)-\frac{f}{2},
\\\label{14}
P_{t}&=&
f_{\mathcal{Q}}\bigg(\frac{\mathcal{Q}}{2}-e^{-\lambda}{\frac{\nu''}{2}
+(\frac{\nu'}{4}+\frac{1}{2r})\times(\nu'-\lambda')}\bigg)-\frac{f}{2},
\\\label{15}
0&=&\frac{\cot\theta}{2}\mathcal{Q}'f_{\mathcal{Q}\mathcal{Q}}.
\end{eqnarray}
In the context of the $f(\mathcal{T})$ theory, the functional form
of $f(\mathcal{T})$ is restricted by the non-zero off-diagonal
metric components of the field equations \cite{94}, which arise from
a particular gauge choice. This limitation also affects the
functional form of $f(\mathcal{Q})$ theory, which is the
mathematical foundation for constructing models based on
$f(\mathcal{Q})$ gravity. In light of the discussion above, we  only
consider $f_{\mathcal{Q}\mathcal{Q}}\neq 0$ in Eq. (\ref{15}) to
obtain the solution of $f(\mathcal{Q})$ gravity and its
corresponding functional form becomes
\begin{equation}\label{16}
f_{\mathcal{Q}\mathcal{Q}}=0\Rightarrow f_{\mathcal{Q}} =
M_{1}\Rightarrow f= M_{1}\mathcal{Q} + M_{2},
\end{equation}
where $M_{1}$ and $M_{2}$ are integration constants. Using
Eq.(\ref{15}), we obtain the field equations as
\begin{eqnarray}\label{17}
\rho&=&\frac{1}{2r^{2}}(2 M_{1}+ 2e^{-\lambda} M_{1}(r
\lambda'-1)-r^{2}M_{2}),
\\\label{18}
P_{r}&=&\frac{1}{r^{2}}(-2 M_{1}+2e^{-\lambda} M_{1}(r
\nu'+1)+r^{2}M_{2}),
\\\label{19}
P_{t}&=&\frac{e^{-\lambda}}{4r}(2e^{\lambda}r M_{2} + M_{1}(2+r
\nu')(\nu'-\lambda')+ 2r M_{1}{\nu''}).
\end{eqnarray}

\subsection{Matching Conditions}

The fundamental boundary metric remains unchanged, whether it is
generated  from the internal and external geometry of the star. This
approach assures that the metric parts will remain continuous over
the boundary surface, irrespective of the coordinate system. The
Jebsen-Birkhoff theorem states that all spherically symmetric vacuum
solutions of field equations have to be asymptotically flat and
static. The Schwarzschild solution is considered as the best option
for the exterior geometry to investigate the CSs. The Schwarzschild
solution can be accommodate for selection for the appropriate
feasible $f(\mathcal{Q})$ gravity model with non-zero density and
pressure. Although, this aspect might cause to violate Birkhoff
theorem \cite{95} in modified theories of gravity. Many researchers
have contributed an immense amount of work on matching criteria.
Goswami et al \cite{99} showed that the junction conditions that
arise in the expanded theories of gravity impose some limitations on
the star objects. For this purpose, many researches
\cite{100}-\cite{104} have considered the Schwarzschild solution,
giving some intriguing outcomes. We consider exterior region as
\begin{equation}\label{20}
ds^{2} = -dt^{2}\aleph+
dr^{2}\aleph^{-1}+d\theta^{2}r^{2}+d\phi^{2}r^{2}\sin^{2}\theta,
\end{equation}
where $\aleph=1-\frac{2M}{r}$. The gravitational collapse of massive
stars generates fascinating objects known as CSs. It is essential to
comprehend their composition, structure and behavior, to improve our
understanding about astrophysics and basic physics. In order to
describe the various aspects and features such as density profiles,
temperature and pressure, we consider the Tolman-Kuchowicz solutions
with arbitrary constants $(\alpha,\beta,B, C)$ as
\begin{equation}\label{21}
e^{\nu} = e^{Br^{2}+2\ln C} , \quad  e^{\lambda}=1+\alpha r^{2}
+\beta r^{4}.
\end{equation}
The unknown constants can be found using the Darmois junction
conditions. By imposing these conditions, researchers can model the
behavior of matter in celestial objects, leading to a deeper
understanding of their physical properties. First Darmois junction
condition governs the matching of solutions across the boundary of
two spacetimes. This condition ensures the continuity of the metric
tensor and its first derivative across the boundary between two
regions of spacetime, where each region can have its own metric
describing the curvature of spacetime. This junction condition is
used to connect two different solutions of field equations, which
describe the gravitational field of a given distribution of matter
and energy. This condition is necessary when there is a boundary,
such as a surface or a hypersurface, between two regions of
spacetime that have different physical properties. This involves
matching the components of the metric tensor and its derivatives at
the boundary, making sure that the gravitational field remains
smooth and continuous across that boundary. By satisfying this
condition, one can create a consistent and continuous spacetime
geometry that smoothly transitions from one solution to another.
This conditions is particularly relevant in scenarios involving
compact objects like black holes or stars, where the gravitational
field transitions from being described by the exterior solution
(e.g., Schwarzschild metric) to an interior solution (e.g., a
solution describing the matter distribution in the object). This is
crucial for constructing physically meaningful spacetimes in such
cases. The second fundamental (continuity of extrinsic curvature at
the hypersurface) gives that radial pressure is zero at the
boundary. Hence, Darmois junction conditions are the mathematical
requirements that enable a coherent description of the gravitational
field in these situations. The continuations of the first and second
fundamental forms at the surface boundary $(r=R)$ gives
\begin{eqnarray}\nonumber
\alpha &=& \frac{1}{R(R-2M)}+\frac{MR}{2(R-2M)^{4}}-\frac{1}{R^{2}},
\\\nonumber
\beta&=&\frac{-M}{2R(R-2M)^{4}}
\\\nonumber
B &=& \frac{M}{R^{2}(R-2M)},
\\\nonumber
C&=&e^{\sqrt{\ln(1-\frac{2M}{R})-\frac{M}{R-2M}}},
\\\nonumber
P_{r}(r=R)&=&0.
\end{eqnarray}
Now, using Tolman-Kuchowicz spacetime constraints resulting filed
equations are
\begin{eqnarray}\label{22}
\rho&=&\frac{-1}{2r^{2}}\bigg[2\alpha r^{2}+6br^{4}+r^{2}-4\bigg],
\\\label{23}
P_{r}&=&\frac{-1}{r^{2}}\bigg[4+4Br^{2}+2\alpha Br^{4}+2\alpha
r^{2}+4B\beta+2\beta r^{4}-r^{2}\bigg],
 \\\nonumber
P_{t}&=& \frac{-(1+\alpha r^{2}+\beta r^{4})}{4r}\bigg[2r(1+\alpha
r^{2}+\beta r^{4})+(2+2Br^{2})
\\\label{24}
&\times&(2Br+2\alpha Br^{3}+2B\beta r^{5}-2\alpha r)(1+\alpha
r^{2}+\beta r^{4})^{-1}+4rB\bigg].
\end{eqnarray}
We consider three CSs models Cen X-3, EXO 1785-248 and LMC X-4. We
used green color for Cen X-3, red color for EXO 1785-248 and blue
color for LMC X-4 for all graphs respectively. The values of mass,
radius and constants for the proposed CSs are given in Table
\textbf{1}.
\begin{table}\caption{Approximated Values of unknown constants.}
\begin{center}
\begin{tabular}{|c|c|c|c|c|c|c|}
\hline Star Models & $\mathrm{M}(km)$ & $\mathrm{R}(km)$ &
$\mathrm{\alpha}  (km^{-2})$ & $\mathrm{\beta}(km^{-2})$  & $B(km^{-2})$ & $C(km^{-2})$ \\
\hline Cen X-3 $\cite{104a}$ & $1.49\pm0.08$ & $9.178\pm0.13$ & 0.0222 & -0.0002 & 0.0030 & 0.7260\\
\hline EXO 1785-248 $\cite{104b}$& $1.30\pm0.2$ &  $10.10\pm0.44$ &  0.0314 & -0.0003 & 0.0039 & 0.7284\\
\hline LMC X-4 $\cite{104a}$ & $1.04\pm0.09$ & $8.301\pm0.2$ & 0.0117 & -0.0006 & 0.0069 & 0.7302\\
\hline
\end{tabular}
\end{center}
\end{table}
In the analysis of stellar objects, it is essential to examine the
behavior of metric elements to ascertain the smoothness and absence
of singularities in the spacetime. The graphical representation
depicted in Figure \textbf{1} serves as a crucial tool in this
evaluation process. It is clear that both metric components display
consistent patterns and show an increasing trend. This behavior is
significant as it indicates the absence of any abrupt or irregular
fluctuations in the spacetime metrics associated with the stellar
objects under consideration. Thus, based on the graphical analysis
presented in Figure \textbf{1}, we can assert that the spacetime
appears to be smooth and devoid of singularities, meeting the
required criteria for our investigation.
\begin{figure}
\epsfig{file=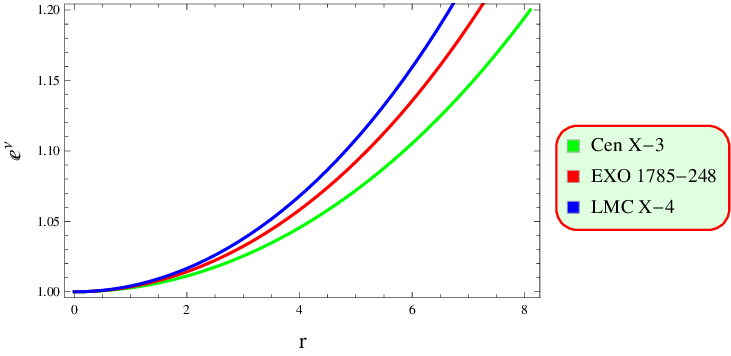, width=.5\linewidth}
\epsfig{file=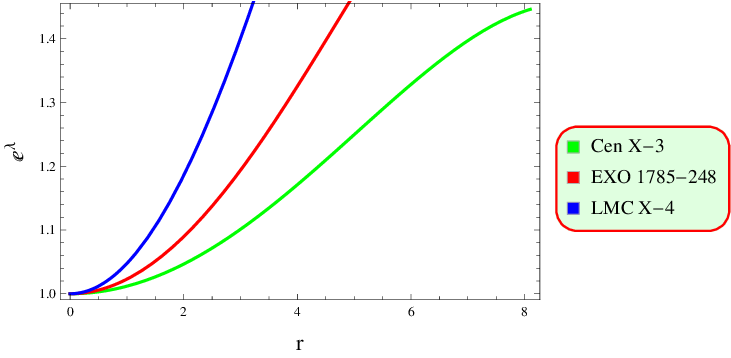,width=.5\linewidth}\caption{Graphical evaluation
of metric potentials corresponding to radial coordinate for
$M_{1}=-0.1$ and $M_{2}=10$.}
\end{figure}

\section{Physical Characteristics}

This section delves into an exploration of the physical
characteristics exhibited by anisotropic CSs, accompanied by
graphical visualizations to examine their characteristics. Our
findings with observational data has the potential to provide
compelling support for the proposed model. Our analysis encompasses
various realistic aspects of compact objects by graphical
representations.

\subsection{Behavior of Matter Contents}

The interior of stars is characterized by an exceptionally dense
profile, leading to the maximum concentration of effective matter
contents. In Figures \textbf{2} and \textbf{3}, we present the
behavior of matter contents and their derivatives for considered
CSs. These figures reveal a positive decrease in the physical
parameters and vanishing points at the boundary, indicating dense
profile of stars. The radial derivative of matter contents exhibit a
consistent pattern, affirming the existence of dense CSs in the
$f(\mathcal{Q})$ framework.

Pressure anisotropy ($\Delta= P_{t} - P_{r}$) refers to the
phenomenon where pressure in a system is not uniform in all
directions. In other words, the pressure can vary depending on the
direction in which it is measured. Positive anisotropy implies
outward pressure, while negative anisotropy signifies inward
pressure. Positive anisotropy refers to the situation where the
tangential pressure $P_{t}$ exceeds the radial pressure $P_{r}$ at a
given point inside the star, i.e., $P_{t} > P_{r}$. This difference
in pressures generates an outward-directed anisotropic force. This
anisotropic pressure can significantly influence the stability and
structure of astrophysical objects, particularly in the modeling of
CSs like neutron stars. In the presence of strong gravitational
forces, particularly near the core of the star, this
outward-directed anisotropy offers an extra counterbalance to
gravity. The outward pressure provided by the positive anisotropy
allows the star to maintain equilibrium at higher densities, which
can potentially lead to a larger maximum mass compared to isotropic
stars. This is significant in explaining the existence of very
massive neutron stars. Positive anisotropy can also result in more
compact stellar configurations, where the star radius is smaller for
a given mass. This compactness may alter the star surface gravity
and its observable properties.

Figure \textbf{4} represents that this anisotropy illustrates an
anti gravitational behavior essential for compact structures.
\begin{figure}
\epsfig{file=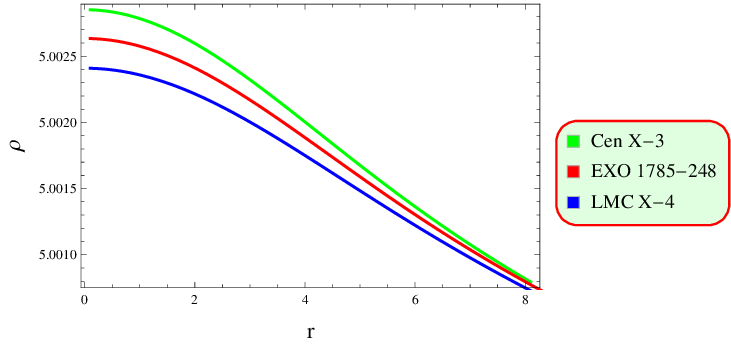,width=.5\linewidth}
\epsfig{file=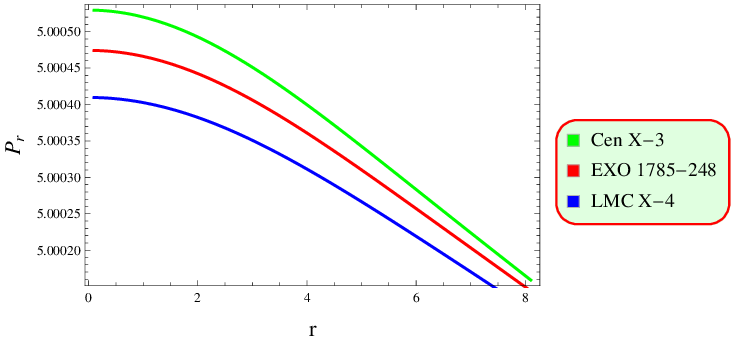,width=.5\linewidth}\center
\epsfig{file=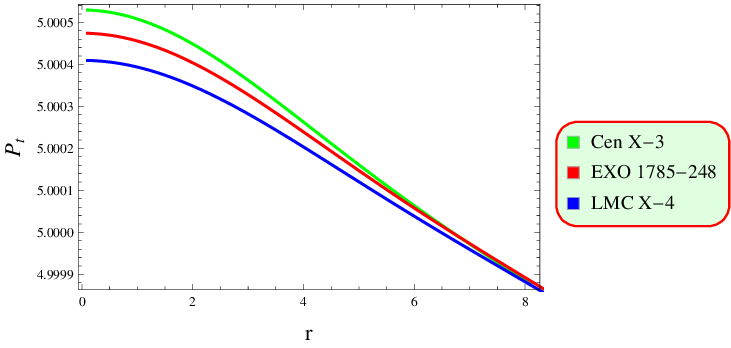,width=.5\linewidth}\caption{Graphs of energy
density, radial pressure and tangential pressure corresponding to
radial coordinate for $M_{1}=-0.1$ and $M_{2}=10$ are maximum at the
center of the stars and monotonically decreasing as the radial
distance increases, revealing a highly compact profile of the
proposed CSs.}
\end{figure}
\begin{figure}
\epsfig{file=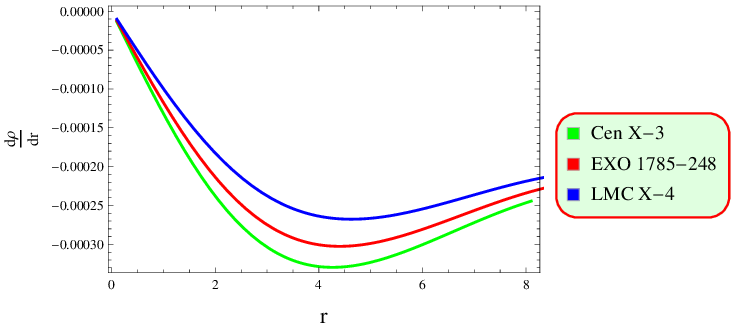,width=.5\linewidth}
\epsfig{file=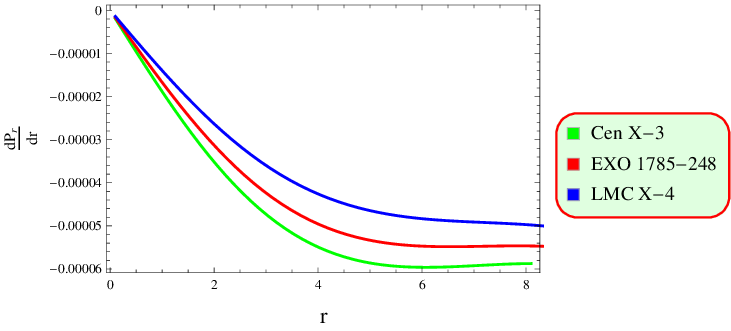,width=.5\linewidth}\center
\epsfig{file=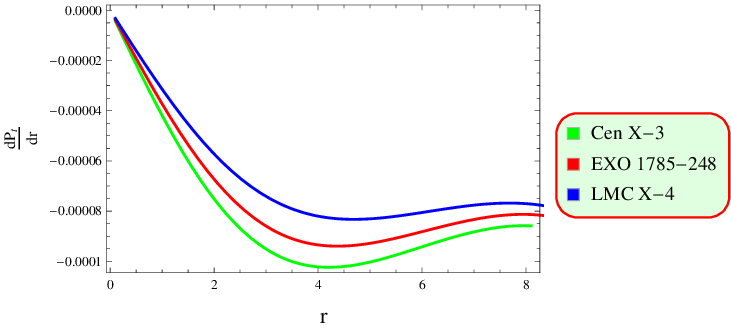,width=.5\linewidth}\caption{Plots of Gradients
of energy density, radial pressure and tangential pressure
corresponding to radial coordinate for $M_{1}=-0.1$ and $M_{2}=10$
are zero at the center and become negative as one moves away from
the core, which confirms the existence of highly compact
configuration in $f(\mathcal{Q})$ theory.}
\end{figure}
\begin{figure}\center
\epsfig{file=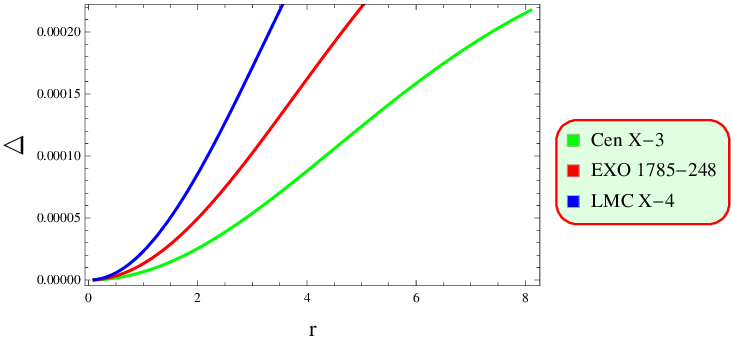,width=.5\linewidth} \caption{Behavior of
anisotropy corresponding to radial coordinate is positively
increasing for all CSs, which ensures the existence of repulsive
force that is necessary for massive geometries.}
\end{figure}

\subsection{Energy Bounds}

The energy conditions consist of inequalities that place constraints
on the stress-energy tensor, governing the interaction between
matter and energy in gravitational field. These constraints are
crucial for assessing the feasibility of CSs, expressed as null
$(9\rho+P_{r}\geq 0, \quad \rho+P_{t}\geq 0)$, dominant $(\rho\pm
P_{r}\geq 0, \quad \rho\pm P_{t}\geq 0)$, weak $(\rho+P_{r}\geq
0,\quad \rho+P_{t}\geq 0, \quad \rho\geq 0)$ and strong
$\rho+P_{r}\geq 0, \quad \rho+P_{t}\geq 0, \quad
\rho+P_{r}+2P_{t}\geq 0$ energy conditions. In Figure \textbf{5}, it
is demonstrated that the considered star candidates maintain
physical viability because they meet all energy constraints in the
presence of $f(\mathcal{Q})$ terms, which affirms the presence of
ordinary matter inside the stars.
\begin{figure}
\epsfig{file=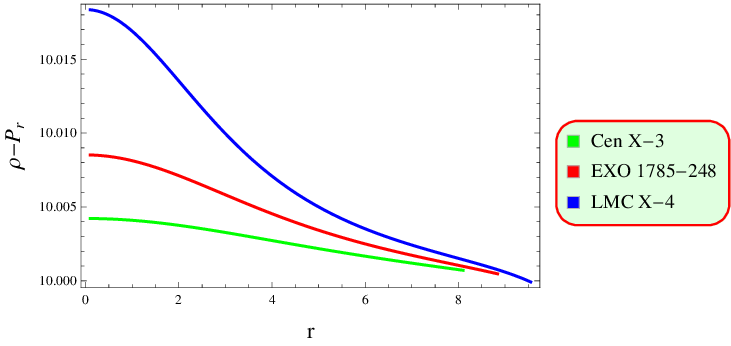,width=.5\linewidth}
\epsfig{file=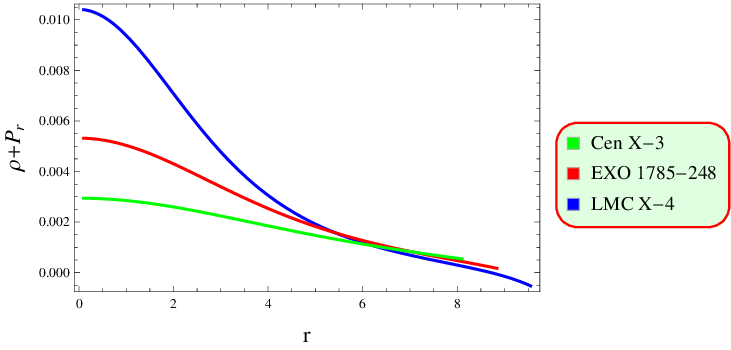,width=.5\linewidth}
\epsfig{file=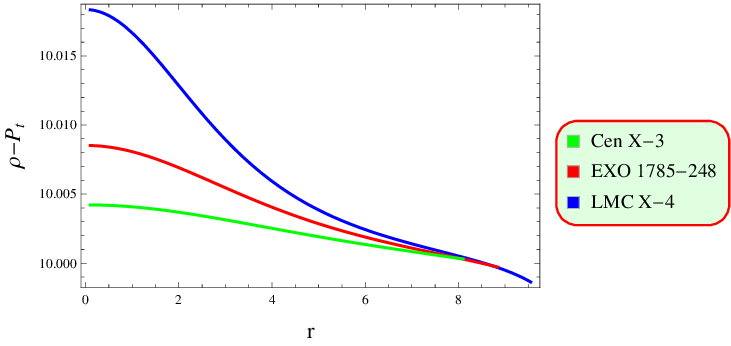,width=.5\linewidth}
\epsfig{file=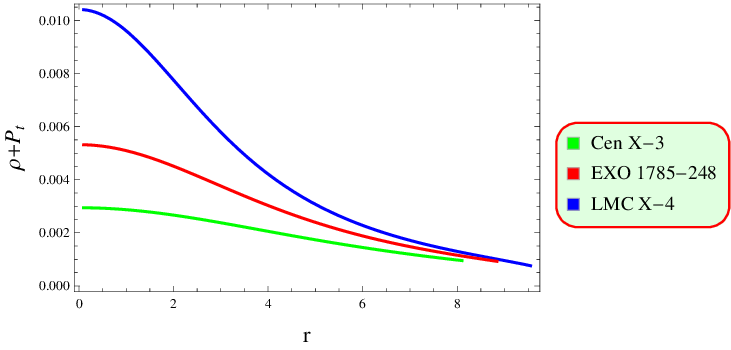,width=.5\linewidth}\center
\epsfig{file=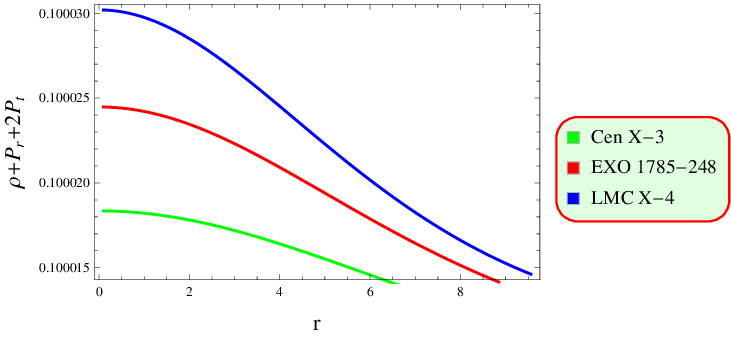,width=.5\linewidth}\caption{Plots of energy
conditions corresponding to radial coordinate for $M_{1}=-0.1$ and
$M_{2}=10$ are positive which demonstrate that the proposed star
candidates are physically viable due to the existence of normal
matter inside the stars.}
\end{figure}

\subsection{Analysis of Different Physical Aspects}

The mass of anisotropic stars is defined as
\begin{equation}\label{25}
M(r) = \int_{0}^{r}\pi r^{2}\rho dr
\end{equation}
In Figure \textbf{6}, the mass function exhibits regularity at the
core of stars under consideration showing a monotonic increase with
radius. This observation indicates the absence of singularities or
irregularities in the mass distribution at the center of these
celestial bodies. To gain insights into the structural composition
of these objects, various physical aspects can be explored. One
crucial parameter for analyzing CSs is the compactness function,
denoted as $(u=\frac{M}{r})$. Buchdahl \cite{105} established a
specific condition for viable CSs, requiring the compactness factor
to satisfy $u<\frac{4}{9}$. This criterion is essential for
understanding the stability of stars.

Another important factor is surface redshift, which measures the
alteration in the wavelength of light emitted from the star surface
as a result of intense gravitational effects. This parameter
provides valuable insight the characteristics of the stars. The
redshift in terms of compactness can be formulated as
\begin{equation}\label{26}
Z_{s}+1 = \sqrt{1-2\mu(r)}
\end{equation}
Figure \textbf{7} shows that surface redshift satisfied the required
limit.
\begin{figure}\center
\epsfig{file=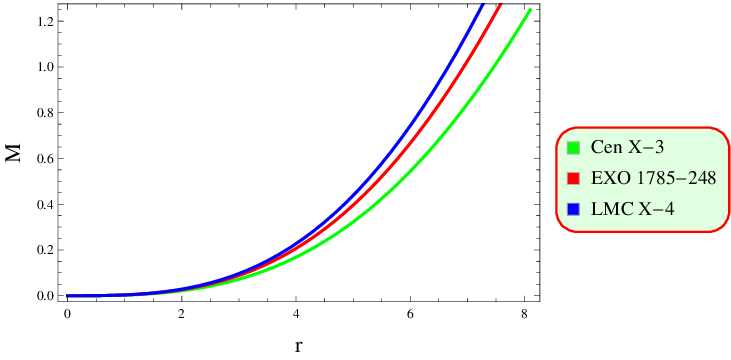,width=.5\linewidth}\caption{Behavior of mass
function corresponding to radial coordinate manifests that the mass
increases positively and monotonically as the radius increases.
Also, $M\rightarrow 0$ as $r\rightarrow 0$ which shows that the mass
function is regular at the center of CSs.}
\end{figure}
\begin{figure}
\epsfig{file=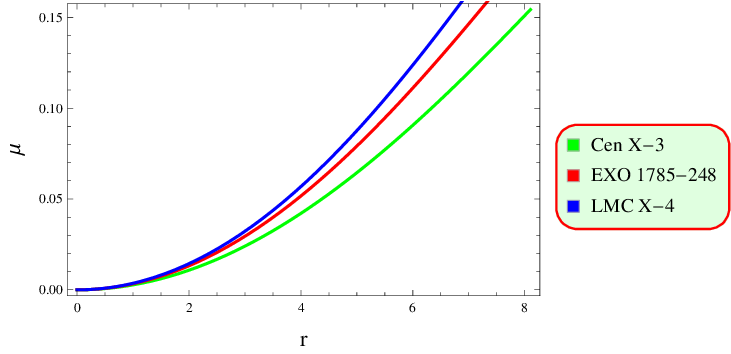,width=.5\linewidth}
\epsfig{file=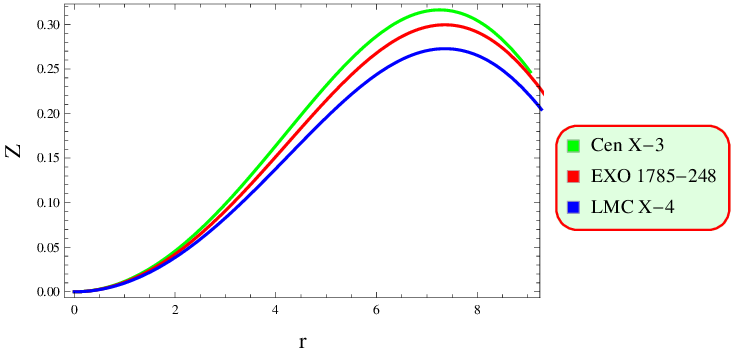,width=.5\linewidth}\caption{Behavior of
compactness and redshift functions corresponding to radial
coordinate lie in the specified limits ($u<4/9$ and $Z_{s}<5.211)$.}
\end{figure}

\subsection{State Parameter}

The equation of state (EoS) parameter plays a crucial role in
understanding the physical properties and internal structure of CSs.
It provides a relationship between the pressure and the energy
density of the matter inside the star, defined as $P=\omega\rho$,
where $\omega$ is the equation of state parameter. The EoS parameter
captures the behavior of nuclear matter, including interactions
between neutrons, protons and possibly hyperons or other exotic
particles. As the density increases towards the star's center,
different phases of matter might emerge and the EoS parameter must
reflect these changes. Thus, the EoS parameter serves as a bridge
between microphysical interactions inside CSs and their macroscopic
observables, providing crucial insights into the nature of
ultra-dense matter and the ultimate fate of massive stars. For
anisotropic CSs, the radial and transversal forms of the EoS
parameters $(\omega_{r}=\frac{P_{r}}{\rho}, \omega_{t} =
\frac{P_{t}}{\rho})$ must lie in [0,1] for the effectiveness of
considered star candidates.
\begin{figure}
\epsfig{file=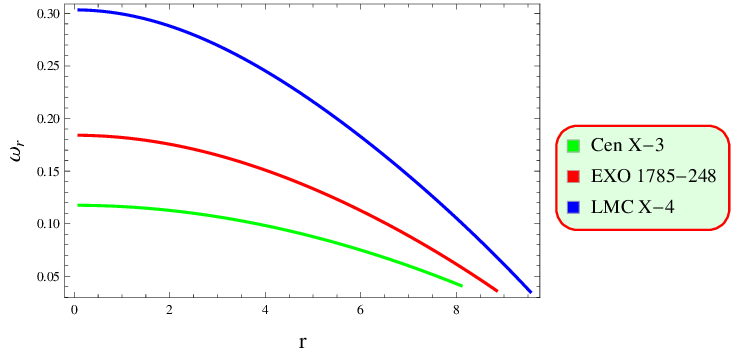,width=.5\linewidth}
\epsfig{file=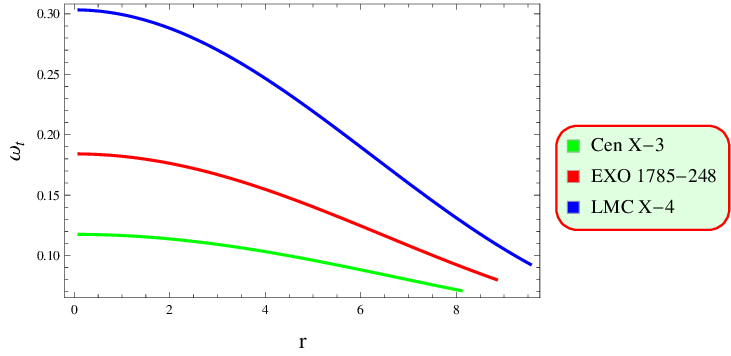,width=.5\linewidth}\caption{Plot of radial and
tangential EoS parameters corresponding to radial coordinate for
$M_{1}=-0.1$ and $M_{2}=10$.}
\end{figure}
The corresponding parameters are described as
\begin{eqnarray}\nonumber
\omega_{r}&=&\frac{-1}{r^{2}}\bigg[4+4Br^{2}+2\alpha Br^{4}+2\alpha
r^{2}+4B\beta+2\beta r^{4}-r^{2}\bigg]
\\\nonumber&\times&
\bigg[\frac{1}{2r^{2}}\{4-2\alpha r^{2}-6br^{4}-r^{2}\}\bigg]^{-1},
\\\nonumber
\omega_{t}&=& \bigg[\frac{-(1+\alpha r^{2}+\beta
r^{4})}{4r}\{2r(1+\alpha r^{2}+\beta
r^{4})\\\nonumber&+&(2+2Br^{2})(\frac{2Br+2\alpha Br^{3}+2B\beta
r^{5}-2\alpha r}{1+\alpha r^{2}+\beta r^{4}})+4rB\}\bigg]
\\\nonumber&\times&
\bigg[\frac{1}{2r^{2}}\{4 - 2\alpha
r^{2}-6br^{4}-r^{2}\}\bigg]^{-1}.
\end{eqnarray}
Figure \textbf{8} shows that our proposed stars are viable as the
state parameters remain in their specified limits.

\section{Stability Analysis}

Stability analysis delves into the repercussions of minor
disturbances on the integrity of stellar objects, investigating
whether they would revert to their initial state of equilibrium or
undergo substantial transformations. It holds significant importance
in gaining insights into the credibility and coherence of cosmic
structures. The examination of conditions ensuring the stability of
these structures against diverse oscillation modes is a key aspect
of stability analysis. The stability is influenced by both the
geometry of these structures and the characteristics of the matter
constituting. Various methods are employed in this context to
analysis the stability of CSs.

\subsection{Tolman-oppenheimer-Volkoff}

This equation provides insight into how different forces acting on a
system maintain its stability. It sheds light on the delicate
balance between a star's internal pressure and the gravitational
force. This equation play pivotal role in gaining insights in
internal composition and characteristics of stars. Now, by using
field equations Eqs.(\ref{17})-Eqs.(\ref{19}), we have TOV equation
in the context of $f(\mathcal{Q})$ theory as follows
\begin{equation}\label{27}
\frac{2\Delta}{r}-\frac{\nu'(\rho+P_{r})}{2}-(P_{r})'=0.
\end{equation}
This equation gives the information on the cosmic balance resulting
from the various forces, i.e., anisotropic $(F_{a})$, hydrostatic
$(F_{h})$ and gravitational $(F_{g})$. These forces in this modified
framework turns out to be
\begin{eqnarray}\nonumber
F_{g}&=& -\frac{B}{2r}\bigg[(4-2\alpha r^{2}-6br^{4}-r^{2})\bigg]+
\frac{1}{r^{2}}\bigg[(4Br^{2}e^{-Ar^{2}}+2e^{-Ar^{2}}+r^{2}-2)\bigg],
\\\nonumber
F_{h}&=&\bigg[\{\frac{-2}{r^{5}}(-4-4Br^{2}-2\alpha Br^{4}-2\alpha
r^{2}-4B\beta-2\beta r^{4}+r^{2})
\\\nonumber&+&\frac{1}{r^{2}}(-8Br-8\alpha Br^{3}-8\alpha r-24B
\beta r^{5}-8\beta r^{3}+2r)\}\bigg],
\\\nonumber
F_{a}&=&-\frac{2}{r}\bigg[ \frac{(1+\alpha r^{2}+\beta
r^{4})}{4r}\{2r(1+\alpha r^{2}+\beta
r^{4})\\\nonumber&+&(2+2Br^{2})(\frac{2Br+2\alpha Br^{3}+2B\beta
r^{5}-2\alpha r}{1+\alpha r^{2}+\beta r^{4}})+4rB\}
\\\nonumber&-&
\{\frac{1}{r^{2}}(-4-4Br^{2}-2\alpha Br^{4}-2\alpha
r^{2}-4B\beta-2\beta r^{4}+r^{2})\}\bigg].
\end{eqnarray}
\begin{figure}
\epsfig{file=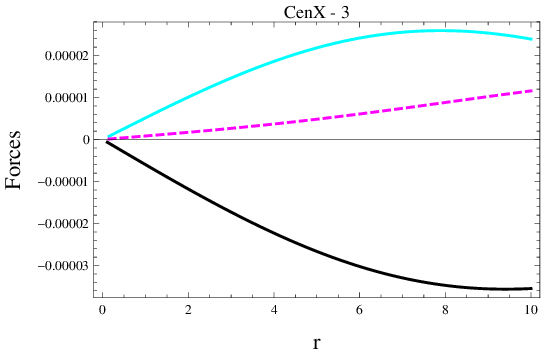,width=.5\linewidth}
\epsfig{file=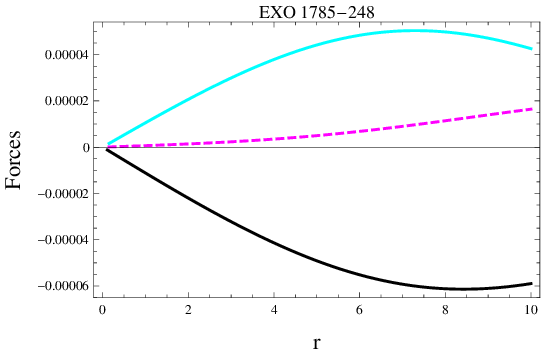,width=.5\linewidth}\center
\epsfig{file=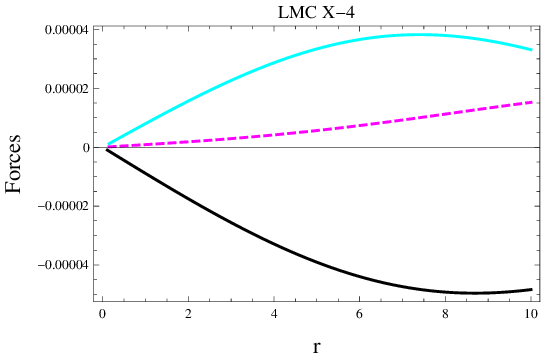,width=.5\linewidth} \caption{ Plot of TOV
equation corresponding to radial coordinate for $M_{1}=-0.1$ and
$M_{2}=10$.}
\end{figure}
Figure \textbf{9} shows that the proposed star candidates are in in
equilibrium condition.

\subsection{Casuality Condition}

We examine the stability through casuality condition for the
considered stars models. In causality condition, the tangential
$\nu_{st} = \frac{dP_{t}}{d\rho}$ and radial $\nu_{sr} =
\frac{dP_{r}}{d\rho}$ components should lie in interval $[0,1]$
\cite{106}. The components of sound speed are given by
\begin{eqnarray}\nonumber
\nu_{sr}&=& \bigg[\{\frac{-2}{r^{5}}(-4-4Br^{2}-2\alpha
Br^{4}-2\alpha r^{2}-4B\beta-2\beta r^{4}+r^{2})
\\\nonumber&+&\frac{1}{r^{2}}(-8Br-8\alpha Br^{3}-8\alpha r-24B\beta r^{5}-8\beta r^{3}+2r)\}\bigg]
\\\nonumber&\times&\bigg[\{\frac{-1}{r^{3}}(4-2\alpha r^{2}-\beta r^{4}-r^{2})+
\frac{1}{2r^{2}}(-4\alpha r-24\beta r^{3}-2r)\}\bigg]^{-1},
\\\nonumber
\nu_{st}&=& \bigg[ \frac{(-1+\alpha r^{3}+\beta
r^{5})}{4r}\{2r(1+\alpha r^{2}+\beta
r^{4})\\\nonumber&+&(2+2Br^{2})(\frac{2Br+2\alpha Br^{3}+2B\beta
r^{5}-2\alpha r}{1+\alpha r^{2}+\beta r^{4}})+4rB\}\bigg]
\\\nonumber&+& \bigg[\{\frac{(1+\alpha r^{2}+\beta
r^{4})}{4r}\}+ \{2+4B+6\alpha r^{2}+10\beta r^{5}\\\nonumber&+&
\frac{8B^{2}r^{2}+8\alpha B^{2}r^{4}+8B^{2}\beta r^{6}-8\alpha
Br^{2}}{1+\alpha r^{2}+\beta r^{4}}+(2+2Br^{2})
\\\nonumber&\times&(\frac{2B+6\alpha Br^{2}}{1+\alpha r^{2}+\beta
r^{4}}-\frac{(2Br+2\alpha Br^{3}+2B\beta r^{5}-2\alpha r)(2\alpha
r+16\beta r^{3})}{(1+\alpha r^{2}+\beta r^{4})^{2}})\}\bigg]
\\\nonumber&\times&\bigg[\{\frac{-1}{r^{3}}(4-2\alpha r^{2}-\beta r^{4}-r^{2})+
\frac{1}{2r^{2}}(-4\alpha r-24\beta r^{3}-2r)\}\bigg]^{-1}
\end{eqnarray}
Figure \textbf{10} determines that the considered stars are stable
and lie in specified limits. Thus, considered stars are physically
stable in $f(\mathcal{Q})$ theory.
\begin{figure}
\epsfig{file=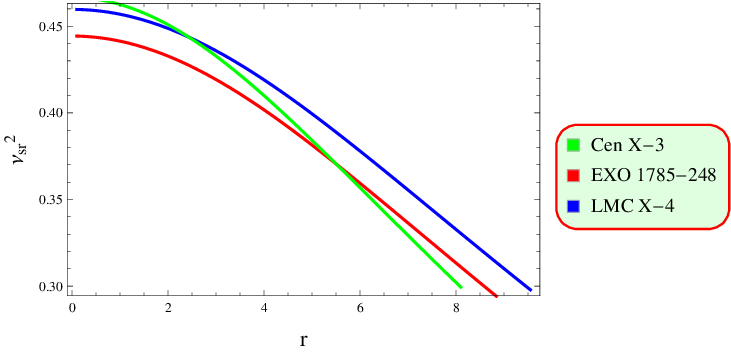,width=.5\linewidth}
\epsfig{file=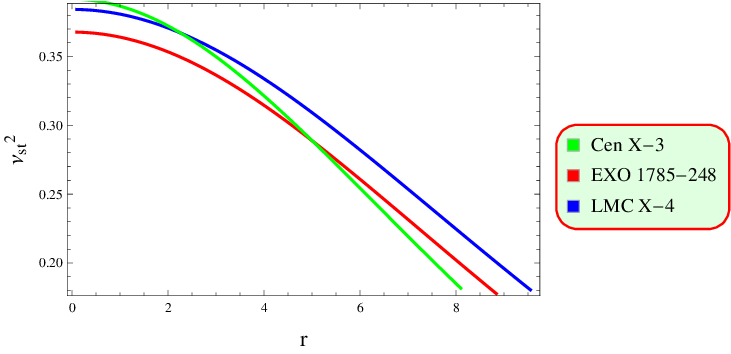,width=.5\linewidth}\caption{Graphs of radial
sound speed component and tangential sound speed component
corresponding to radial coordinate lie in [0,1] in the presence of
modified terms, which show that spherically symmetric solutions are
in the stable state.}
\end{figure}

\subsection{ Herrera Cracking Approach}

The concept of cracking is associated to the tendency of a fluid
distribution to ``split'', once it abandons the equilibrium as
consequence of perturbations. Thus, one can say that once the system
has abandoned the equilibrium, there is a cracking, whenever its
inner part tends to collapse whereas its outer part tends to expand.
The cracking takes place at the surface separating the two regions.
When the inner part tends to expand and the outer one tends to
collapse we say that there is an overturning. It is worthwhile to
mention that the concepts of stability and cracking are different,
although they are often confused. The term stability refers to the
capacity of a given fluid distribution to return to equilibrium once
it has been removed from it. The fact that the speeds of sound are
not superluminal does not assure in any way the stability of the
object, it only ensures causality. The cracking only implies the
tendency of the system to split immediately after leaving the
equilibrium. Whatever happens next, whether the system enters into a
dynamic regime, or returns to equilibrium, is independent of the
concept of cracking. Of course the occurrence of cracking will
affect the future of the fluid configuration in either case.

We use a Herrera cracking method for the stability of stars
\cite{107}. A crucial factor in identifying the stability regions is
difference between sound speed in radial and transverse directions.
The area is considered unstable if the inequality is not satisfied
by this difference. In particular for stable regions,
$0\leq|\nu_{st}-\nu_{sr}|\leq 1$. The stars under consideration,
satisfy these requirements as shown in Figure \textbf{11} which
implies a stable state.
\begin{figure}\center
\epsfig{file=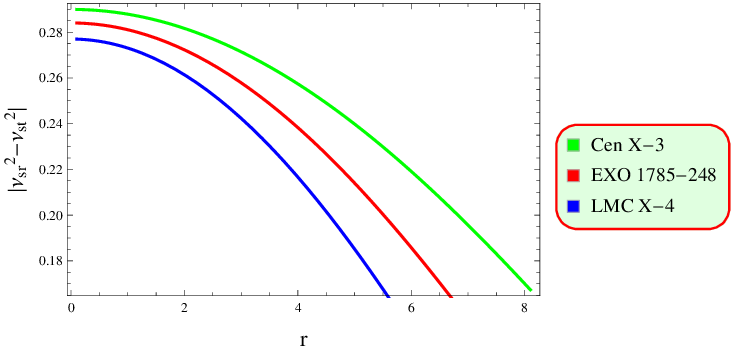,width=.5\linewidth}\caption{Graphs of Herrera
cracking corresponding to radial coordinate determine that
considered CSs are stable as they lie in the specified limit.}
\end{figure}

\subsection{Adiabatic Index}

Adiabatic index is another method which is used for the stability of
CSs. The radial and tangential components are given as
\begin{eqnarray}\label{28}
\Gamma_{r}=(\frac{P_{r}+\rho}{P_{r}})\nu_{sr}, \quad \Gamma_{t}
=(\frac{P_{t}+\rho}{P_{t}})\nu_{st}.
\end{eqnarray}
Using Eqs.(\ref{23})-Eqs.(\ref{25}), we have
\begin{eqnarray}\nonumber
\Gamma_{r}&=&  \bigg[\{\frac{1}{2r^{2}}(4 - 2\alpha
r^{2}-6br^{4}-r^{2})\}+\{\frac{1}{r^{2}}(-4-4Br^{2}\\\nonumber&-&2\alpha
Br^{4}-2\alpha r^{2}-4B\beta-2\beta
r^{4}+r^{2})\}\bigg]\\\nonumber&\times&\bigg[\frac{1}{r^{2}}(-4-4Br^{2}-2\alpha
Br^{4}-2\alpha r^{2}-4B\beta-2\beta r^{4}+r^{2})\bigg]^{-1}
\\\nonumber&\times&\bigg[\{\frac{-2}{r^{5}}(-4-4Br^{2}-2\alpha
Br^{4}-2\alpha r^{2}-4B\beta-2\beta r^{4}+r^{2})
\\\nonumber&+&\frac{1}{r^{2}}(-8Br-8\alpha Br^{3}-8\alpha r-24B\beta r^{5}-8\beta r^{3}+2r)\}\bigg]
\\\nonumber&\times&\bigg[\{\frac{-1}{r^{3}}(4-2\alpha r^{2}-\beta r^{4}-r^{2})+
\frac{1}{2r^{2}}(-4\alpha r-24\beta r^{3}-2r)\}\bigg]^{-1}
\\\nonumber\Gamma_{t}&=&\bigg[\{\frac{1}{2r^{2}}(4 - 2\alpha
r^{2}-6br^{4}-r^{2})\}+\{\frac{-(1+\alpha r^{2}+\beta
r^{4})}{4r}(2r(1+\alpha r^{2}+\beta
r^{4}))\\\nonumber&+&(2+2Br^{2})(\frac{2Br+2\alpha Br^{3}+2B\beta
r^{5}-2\alpha r}{1+\alpha r^{2}+\beta r^{4}})+4rB\bigg]
\\\nonumber
&\times&\bigg[\frac{-(1+\alpha r^{2}+\beta r^{4})}{4r}\{2r(1+\alpha
r^{2}+\beta r^{4})\\\nonumber&+&(2+2Br^{2})(\frac{2Br+2\alpha
Br^{3}+2B\beta r^{5}-2\alpha r}{1+\alpha r^{2}+\beta
r^{4}})+4rB\}\bigg]^{-1}
\\\nonumber&\times&
\bigg[ \frac{(-1+\alpha r^{3}+\beta r^{5})}{4r}\{2r(1+\alpha
r^{2}+\beta r^{4})\\\nonumber&+&(2+2Br^{2})(\frac{2Br+2\alpha
Br^{3}+2B\beta r^{5}-2\alpha r}{1+\alpha r^{2}+\beta
r^{4}})+4rB\}\bigg]
\\\nonumber&+& \bigg[\{\frac{(1+\alpha r^{2}+\beta
r^{4})}{4r}\}+ \{2+4B+6\alpha r^{2}+10\beta r^{5}\\\nonumber&+&
\frac{8B^{2}r^{2}+8\alpha B^{2}r^{4}+8B^{2}\beta r^{6}-8\alpha
Br^{2}}{1+\alpha r^{2}+\beta r^{4}}+(2+2Br^{2})
\\\nonumber&\times&(\frac{2B+6\alpha Br^{2}}{1+\alpha r^{2}+\beta
r^{4}}-\frac{(2Br+2\alpha Br^{3}+2B\beta r^{5}-2\alpha r)(2\alpha
r+16\beta r^{3})}{(1+\alpha r^{2}+\beta r^{4})^{2}})\}\bigg]
\\\nonumber&\times&\bigg[\{\frac{-1}{r^{3}}(4-2\alpha r^{2}-\beta r^{4}-r^{2})+
\frac{1}{2r^{2}}(-4\alpha r-24\beta r^{3}-2r)\}\bigg]^{-1}
\end{eqnarray}
Chandrasekhar determine the condition that system is stable if
$\Gamma > \frac{4}{3}$ otherwise unstable \cite{108}. For all
considered CSs, our system is stable as shown in Figure \textbf{12}.
\begin{figure}
\epsfig{file=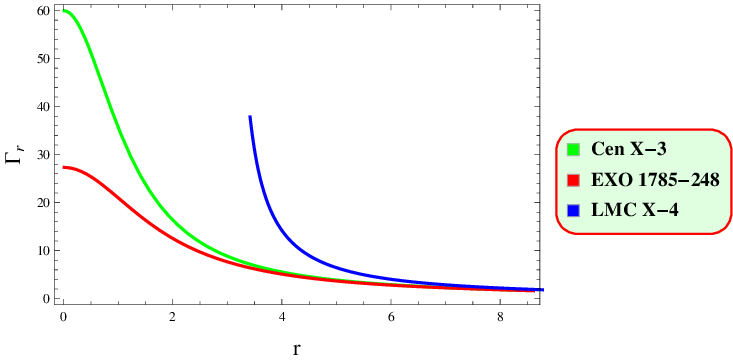,width=.5\linewidth}
\epsfig{file=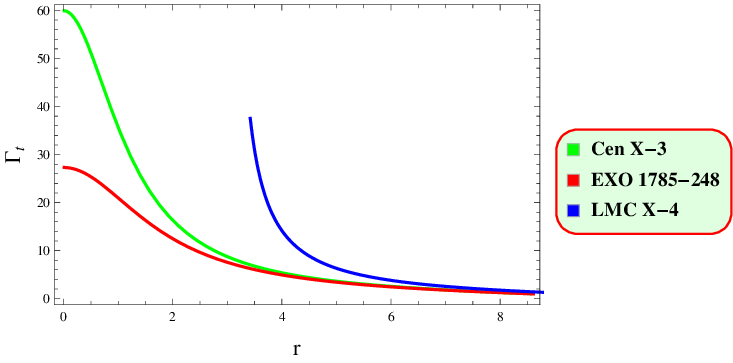,width=.5\linewidth}\caption{Plot of adiabatic
index corresponding to radial coordinate for $M_{1}=-0.1$ and
$M_{2}=10$.}
\end{figure}

\section{Conclusions}

Modified gravity theories such as $f(\mathcal{Q})$ gravity have
received considerable interest in recent years as alternate
explanations to GR. This theory suggests alterations to the
fundamental equations of gravity with the aim of addressing numerous
unresolved inquiries in the fields of cosmology and astrophysics. A
crucial method for assessing and limiting the validity of this
modified gravity theory is by conducting astrophysical observations.
This modified theory introduces extra independent variables that
describe deviations from GR. This modified proposal aims to enhance
our understanding of the principles that regulate the actions of
matter and energy in the cosmos, providing opportunities to
investigate novel phenomena that may not be accounted for in GR. The
inclusion of correction terms emphasize the significant results,
imposing a large impact on the geometric feasibility. The
$f(\mathcal{Q})$ theory is used as a mathematical tool to examine
complex elements of gravitational dynamics on a large scale. The
motivation behind the investigation of this theory involves
analyzing its theoretical implications, its consistency with
empirical facts and its importance in cosmological contexts.

In recent years, the discipline of theoretical physics has been
greatly interested in studying CSs. These celestial entities provide
considerable obstacles to our understanding of basic physics and
offer distinct possibilities for studying extreme situations in the
theoretical framework. Our study is centered around the
investigation of anisotropic CSs in $f(\mathcal{Q})$ theory with the
objective of exploring the enigmatic aspects of the cosmos. By
examining anisotropic CSs in this theoretical framework,
gravitational interactions unveil on both galactic and cosmic
levels. This offers valuable understanding of how these components
impact stellar structures. Studying these objects in this theory
allows us to investigate the behavior of gravity under conditions of
high curvature and density as gravity in CSs approaches its extreme
limits. By examining the behavior of these compact celestial
objects, we acquire significant knowledge about the properties of
CSs.

This paper delves into the examination of the viability and
stability of CSs utilizing the Tolman-Kuchowicz solution in the
framework of $f(\mathcal{Q})$ theory. The values of unknown
constants are given in Table \textbf{1}. The specific model of this
modified theory is developed to investigate the behavior of
different physical quantities within the inner regions of stars. The
key findings are summarized below.
\begin{itemize}
\item
We have found that both metric elements (Figure \textbf{1}) are
consistent and fulfill the necessary conditions, i.e., they exhibit
minimum value at the center of stars and then show monotonically
increasing behavior.
\item
The behavior of fluid parameters (Figure \textbf{2}) is positive and
regular in the interior of CSs and diminish at the boundary. Also,
the derivative of fluid parameters (Figure \textbf{3}) is negative
which presents a dense picture of the CSs.
\item
We have found that the anisotropic pressure (Figure \textbf{4}) is
directed outward which is necessary for compact stellar
configuration
\item
All energy bounds are satisfied to confirm the presence of normal
matter in the interior of CSs (Figure \textbf{5}).
\item
We have found that the mass function is regular at the center of the
star and show monotonically increasing behavior as the radial
coordinate increases Figure \textbf{6}. The compactness and redshift
functions satisfy the required conditions Figure \textbf{7}.
\item
The range of EoS parameters (Figure \textbf{8}) lies between 0 and
1, which shows the viability of the considered model.
\item
The TOV equation shows that gravitational, hydrostatic, and
anisotropic forces have a null impact for all proposed CSs (Figure
\textbf{}). This suggests that the compact stellar models are in an
equilibrium state.
\item
The stability limits, i.e., $\nu_{sr}$ and $\nu_{st}$ lies between 0
and 1 (causality condition), $0<| \nu^{2}_{t}-\nu^{2}_{r}|<1$
(Herrera cracking) approach and $\Gamma>\frac{4}{3}$ (adiabatic
index) are satisfied, which ensures the existence of physically
stable CSs (Figures \textbf{10}-\textbf{12}).
\end{itemize}
We have conducted an investigation into the viability and stability
of CSs incorporating $f(\mathcal{Q})$ terms. Our analysis has
unveiled a denser profile for CSs, demonstrated through a
comprehensive examination of the derived solutions. In the context
of $f(\mathcal{R},\mathcal{T}^{2})$ theory, it has been established
that CSs are not physically feasible or stable at their center
\cite{80}. Our findings indicate that all the considered CSs exhibit
both physical feasible as well as stability at their central
position in $f(\mathcal{Q})$ theory. Therefore, our derived
solutions provide stable formations of anisotropic CSs.

\end{document}